\documentclass[useAMS,usenatbib,usegraphicx]{mn2e}

\title[Influence of AGN Outbursts]{Influence of AGN Outbursts
on the Surrounding Galaxies}
\author[Y. Fujita]{Yutaka Fujita\thanks{E-mail:
fujita@vega.ess.sci.osaka-u.ac.jp}\\
Department of Earth and Space Science, Graduate School
of Science, Osaka University, \\
1-1 Machikaneyama-cho, Toyonaka, Osaka
560-0043, Japan}
\begin{document}

\date{Accepted 2007 December 15. Received 2007 December 14; in original form 2007 October 11}

\pagerange{\pageref{firstpage}--\pageref{lastpage}} \pubyear{2007}

\maketitle

\label{firstpage}

\begin{abstract}
 We study the influence of a strong AGN outburst on the surrounding
 galaxies. The AGN is assumed to reside in a group of galaxies, and an
 outburst excites a shock wave in the hot gas in the group. We calculate
 the impact of the shock wave on the galaxies. We find that if the
 energy of the outburst is extremely large ($E_{\rm AGN}\sim 6\times
 10^{61}$~erg) as the one recently observed in clusters, the impact is
 strong enough to strip the cold interstellar medium in the disc of the
 galaxies in the inner region of the group. Moreover, even in the outer
 region of the group, the warm gas in the halo of the galaxies would be
 stripped, even if the energy of the outburst is $\sim 6\times
 10^{60}$~erg. These would decrease star formation activity of the
 galaxies. If these galaxies fall into the group centre through
 dynamical friction and their interstellar medium is the fuel of the
 supermassive black hole in the AGN, the outburst would serve as
 feedback. While this mechanism works only when $E_{\rm AGN}$ is
 extremely large, such outbursts have not been observed in groups at low
 redshift; it would work at high redshift rather than at low redshift.
\end{abstract}

\begin{keywords}
galaxies: active -- galaxies: clusters: general -- galaxies:
interactions -- galaxies: intergalactic medium.
\end{keywords}

\section{Introduction}
\label{sec:intro}

X-ray observations have shown that hot gas in groups and clusters of
galaxies has been heated by some sources in addition to gravity. This
was shown by the fact that the luminosity and temperature of a group or
cluster follow a scaling relationship (the $L_\rmn{X}$--$T$ relation),
$L_\rmn{X}\propto T^3$, \citep{edg91,all98,mar98,arn99}, which is at
odds with that expected for groups and clusters formed by gravitational
structure formation, with $L_\rmn{X}\propto T^2$ \citep{kai86}. More
recently, it was shown that the entropies of the hot gas in groups and
clusters, especially groups, are higher than those predicted by models of
gravitational structure formation \citep*{pon99}.

Supernova-driven galactic winds have been considered the heating source
\citep*{wu98,men00,loe00}. However, the energy from supernovae alone
seems to be insufficient to heat the hot gas to the observed level
\citep*{val99,wu00,bow01}. Therefore, active galactic nuclei (AGNs) are
now recognised as another promising candidate of the heating source
\citep{ino01}.

AGNs are often found at the centres of groups and clusters
\citep{mcn00,fab00,bla01}. The energy ejected by an AGN can cancel
radiative cooing of the hot gas and may prevent development of a cooling
flow in the central region of a group or a cluster. For most of the AGNs
at $z\sim 0$, however, the power is not enough to heat the hot gas on a
group or cluster-scale, and would be insufficient to account for the
entropy excess found in groups and clusters. Recently, however, shock
waves associated with extremely powerful AGN outbursts have been found
in some clusters \citep*{mcn05,nul05a,nul05b,wis07,git07}. They seem to
be powerful enough to heat the gas on a group or cluster-scale. 

Such strong outbursts would also affect the surrounding environment in a
form other than heating. For example, \citet{fuj07} showed that the
shock wave excited by an outburst accelerates particles and that the
emission from the accelerated particles could be responsible for 
radio mini-halos observed in clusters. \citet{raw04} discussed the
regulation of galaxy formation by outbursts.

In this letter, we consider another effect of strong AGN outbursts on
the environment. We focus on the interaction between the shock wave
produced by an outburst and the galaxies surrounding the AGN. We
consider an AGN and galaxies in a small group of galaxies with the mass
of $10^{13}\rm\: M_\odot$ rather than a cluster of galaxies. In
clusters, another interaction between the hot gas and the galaxies,
called ram-pressure stripping, is effective (see
Section~\ref{sec:discuss}). Thus, the influence of AGN outbursts would
be obscured.

It should be noted that such strong outbursts are rare phenomena at low
redshift. \citet{git07} estimated that strong outbursts ($E_{\rm
AGN}\sim 10^{61}$~erg) are likely to occur only $\sim 10$\% of clusters
at low redshift. For groups, the fraction may even be smaller because
such outbursts have not been observed. However, at high redshift, they
would be more common \citep[][see Section~\ref{sec:discuss}]{ued03}.  In
this letter, the cosmological parameters are $\Omega_0=0.3$,
$\lambda_0=0.7$, and $h=0.7$, where $H_0=100\: h\rm\: km\: s^{-1}\:
Mpc^{-1}$.

\section[]{Models}
\subsection{Dark Matter and Gas Profile}

Using one-dimensional hydrodynamic simulations, we estimate the impact
of the shock wave created by an AGN outburst on galaxies. We assume
that an AGN is located at the centre of a group of galaxies with the
virial mass of $M_\rmn{vir}$ and that the group is spherically
symmetric.  For the mass distribution of the group, we adopt the
so-called NFW profile \citep*{nav97}, although later studies indicated
that the central cusp would be steeper \citep[e.g.][]{fuk07}. The mass
profile is written as
\begin{equation}
\label{eq:NFW}
 M(R) \propto \left[\ln \left(1+\frac{R}{R_s}\right)
-\frac{R}{R_s (1+R/R_s)}
\right]\:,
\end{equation}
where $R_s$ is the characteristic radius of the group.  The
normalisation can be given by $M(R_{\rm vir})=M_{\rm vir}$, where
$R_{\rm vir}$ is the virial radius of the group.

The virial radius is given by
\begin{equation}
 R_\rmn{vir}=\left[\frac{3M_\rmn{vir}}
{4\pi\Delta_c(z)\rho_c(z)}\right]^{1/3}\:,
\end{equation}
where $\Delta_c(z)$ is a spherical over-density of the virialized dark
halo within $R_\rmn{vir}$ at redshift $z$, in units of the critical
density of the Universe at $z$, or $\rho_c(z)$. For $\Delta_c(z)$, we
use the fitting formula of \citet{bry98} for a flat Universe with a
non-zero cosmological constant, $\Delta_c(z)=18\pi^2 +82x -39x^2$,
where $x=\Omega(z)-1$ and $\Omega(z)$ is the cosmological density
parameter at redshift $z$. The concentration parameter of the group,
$c_\rmn{vir}=R_\rmn{vir}/R_\rmn{s}$, is given by
\begin{equation}
 c_\rmn{vir}=\frac{9}{1+z}
\left(\frac{M_\rmn{vir}}
{1.5\times 10^{13} h^{-1}\rmn{M_\odot}}\right)^{-0.13}
\end{equation}
\citep{bul01}.

Initially, hot gas or intragroup medium (IGM) is in pressure equilibrium
with the gravitational potential formed by the group. We assume that the
initial IGM density and temperature profiles follow the `universal
profile' derived by \citet{kom01}. They can respectively be written as
\begin{equation}
 \rho_\rmn{IGM}(R)=\rho_\rmn{IGM}(0)y_\rmn{IGM}(R/R_\rmn{s})\:,
\end{equation}
\begin{equation}
 T_\rmn{IGM}(r)=T_\rmn{IGM}(0)y_\rmn{IGM}^{\gamma'-1}(R/R_\rmn{s})\:,
\end{equation}
where
\begin{equation}
 y_\rmn{IGM}^{\gamma'-1}(x)=1-\frac{3}{\eta_0}\frac{\gamma'-1}{\gamma'}
\frac{c_\rmn{vir}}{m(c_\rmn{vir})}
\left[1-\frac{\ln(1+x)}{x}\right]\:,
\end{equation}
\begin{equation}
 m(x)=\ln(1+x)-x/(1+x)\:.
\end{equation}
The parameters $\eta_0$ and $\gamma'$ can be derived from the condition
that the IGM and dark matter profiles are the same in the outermost
region of the group \citep{kom01}. 

We solve the following equations:
\begin{equation}
 \frac{\partial \rho_\rmn{IGM}}{\partial t}
+ \frac{1}{R^2}\frac{\partial}{\partial R}
(R^2 \rho_\rmn{IGM} V_\rmn{IGM})=0 \;,
\end{equation}
\begin{eqnarray}
 \frac{\partial (\rho_\rmn{IGM} V_\rmn{IGM})}{\partial t}
+\frac{1}{R^2}\frac{\partial}{\partial R}
(R^2 \rho_\rmn{IGM} V_\rmn{IGM}^2)\nonumber\\ 
=-\rho_\rmn{IGM} \frac{G M(R)}{R^2}-\frac{\partial p}{\partial R}
\;,
\end{eqnarray}
\begin{eqnarray}
\label{eq:energy2}
 \frac{\partial e}{\partial t}
+\frac{1}{R^2}\frac{\partial}{\partial R}[R^2 V_\rmn{IGM}(p+e)]
=-n_\rmn{e}^2 \Lambda(T_\rmn{IGM})
\nonumber\\
-\rho_\rmn{IGM} V_\rmn{IGM} G M(R)/R^2 \;,
\end{eqnarray}
where $G$ is the gravitational constant, and $p$ and $V_\rmn{IGM}$ are
the pressure and velocity of the IGM, respectively. The total energy is
defined as $e=p/(\gamma-1)+\rho_\rmn{IGM} V_\rmn{IGM}^2/2$, where
$\gamma=5/3$. Although we include the cooling function
$\Lambda(T_\rmn{IGM})$, the evolution of the shock is faster than the
radiative cooling. The electron density is defined as
$n_\rmn{e}=0.86\:\rho_\rmn{IGM}/m_\rmn{p}$, where $m_\rmn{p}$ is the
proton mass. We ignore the self-gravity of IGM.

\subsection{The Criterion for Stripping}

The cold interstellar medium (ISM) in the disc of a galaxy would be
stripped when a shock wave excited in the IGM passes the galaxy. Since
the shock wave passes the galaxy in a short time (less than the
dynamical time or the rotation time of the galaxy), momentum transfer
causes stripping to occur. The integrated momentum from the IGM per
unit area at radius $R$ is
\begin{equation}
\label{eq:mom}
s(t,R)= \int_0^t \rho_\rmn{IGM}(t,R)V_\rmn{IGM}^2(t,R)dt\;.
\end{equation}
Thus, the criterion of stripping for a galaxy at radius $R$ is
\begin{equation}
\label{eq:criterion}
s(t,R)>\Sigma_\rmn{ISM}v_\rmn{esc}\;,
\end{equation}
where $\Sigma_\rmn{ISM}$ is the column density of the cold ISM in the
galactic disc, and $v_\rmn{esc}$ is the escape velocity of the
galaxy. Strictly speaking, this relation is valid only when the galaxy
is face-on; if not, the stripping would be less efficient. In this
relation, we assume that the galaxy is not moving relative to the
group. The AGN explodes at $t=0$.

The escape velocity is given by $v_\rmn{esc}\sim
\sqrt{2}v_\rmn{rot}$, where $v_\rmn{rot}$ is the rotation velocity of
the galaxy. (The ISM may stay in the halo of the galaxy with this
velocity.) \citet*{mo98} indicated that for a given $v_\rmn{rot}$ the
total disc surface density of a galaxy, $\Sigma_\star$, is proportional
to the Hubble constant at redshift $z$:
\begin{equation}
 H(z)=H_0[\lambda_0+(1-\lambda_0-\Omega_0)(1+z)^2 + \Omega_0(1+z)^3]^{1/2}
\end{equation}
Following \citet{fuj04}, we assume that the ISM column density is
proportional to the disc surface density ($\Sigma_\rmn{ISM}\propto
\Sigma_\star$). Thus, the former is given by
\begin{equation}
\label{eq:Sig_ISM}
 \Sigma_\rmn{ISM}(z)=\Sigma_\rmn{ISM}(0)H(z)/H_0 \:.
\end{equation}
Moreover, we assume that the disc radius of a galaxy has a relation of
\begin{equation}
\label{eq:r_gal}
 r_\rmn{gal}(z)=r_\rmn{gal}(0)[H(z)/H_0]^{-1}\:
\end{equation}
\citep{mo98}.

\section{Results}
\label{sec:result}

We consider the influence on galaxies in groups exerted by an extremely
strong AGN outburst that has not been observed in low-redshift groups.
We assume that the mass of a galaxy group is $M_\rmn{vir}=1\times
10^{13}\:\rm M_\odot$. We set $\Sigma_\rmn{ISM}(0)=8\times 10^{20}
m_\rmn{p}$ and $r_\rmn{gal}(0)=10$~kpc. We fix the rotation and escape
velocities of the galaxy at $v_\rmn{rot}=220\rm\; km\: s^{-1}$ and
$v_\rmn{esc}=\sqrt{2} v_\rmn{rot}$, respectively. These parameters are
those for the Galaxy \citep[e.g.][]{spi78}.

Since we do not know much about the sources of the non-gravitational
heating in groups and clusters (Section~\ref{sec:intro}), we assume that
the IGM has not been non-gravitationally heated at $t=0$ for the sake of
simplicity. In other words, the IGM is non-gravitationally heated for
the first time by the AGN outburst we consider below. Thus, we assume
that the mass fraction of the IGM in the group is the same as the baryon
fraction of the Universe and is 0.15. If the IGM has been
non-gravitationally heated, the mass fraction and density of the IGM
would be lower and the influence of the shock wave on galaxies would be
smaller.

First, we assume that the AGN at the group centre ejects an energy of
$E_{\rm AGN}=6\times 10^{61}$~erg, which is the one estimated for the
cluster MS~0735.6$+$7421 \citep{mcn05}. The energy is kinematically
given to the IGM for $R=10$--20~kpc at $t=0$; the details of the energy
input do not affect the results. We use 1000 unequally spaced meshes in
the radial coordinate to cover a region with a radius of 600~kpc.  The
inner boundary is set at $R=1$~kpc.

\begin{figure}
\includegraphics[width=74mm]{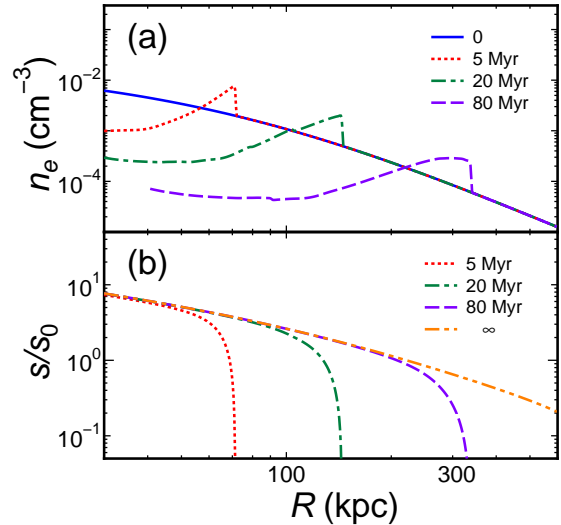} \caption{The evolutions of profiles
 at $z=0$. (a) The density profiles at $t=0$, 5, 20, and 80~Myr. The
 curve at $t=80$~Myr is omitted at small $R$ because it is affected by
 the inner boundary condition. (b) The profiles of $s/s_0$ at $t=5$, 20,
 80~Myr, and $t=\infty$.}  \label{fig:z0}
\end{figure}

Fig.~\ref{fig:z0}(a) shows the evolution of the IGM density profile for
the group at $z=0$. The parameters of the group are
$R_\rmn{vir}=560$~kpc and $c_\rmn{vir}=9.9$. The outburst is strong
enough to blow away most of the IGM. The shock reaches the virial radius
at $t\approx 1.6\times 10^8$~yr. Fig.~\ref{fig:z0}(b) shows the
evolution of $s/s_0$, where $s_0=\Sigma_\rmn{ISM}v_\rmn{esc}$ (see
equation~\ref{eq:criterion}). Fig.~\ref{fig:z0}(b) indicates that the
cold ISM in the galactic disc is stripped for $R<220$~kpc because
$s/s_0>1$ at $t=\infty$. The profile $s(t,R)$ does not change after the
shock passes the outermost region of the group. We refer to this final
profile as $s_\rmn{f}(R)$ [$=s(t=\infty,R)$]. Of course, the ISM of
galaxies with $v_\rmn{rot}<220\rm\: km\: s^{-1}$ is more easily
stripped for a given $\Sigma_\rmn{ISM}$.

\begin{figure}
\includegraphics[width=74mm]{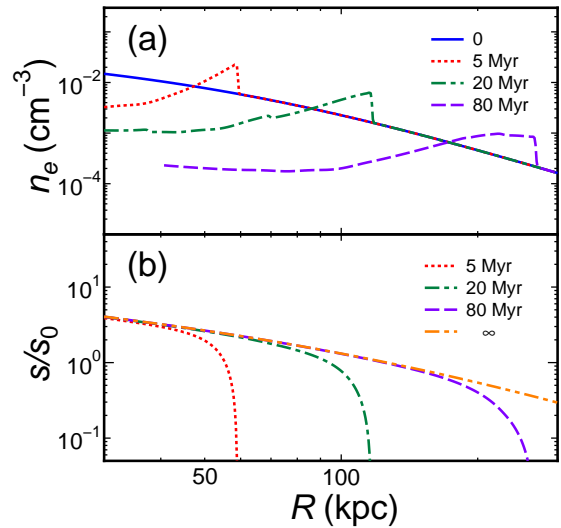} \caption{Same as Fig.~\ref{fig:z0}
 but for $z=2$.}  \label{fig:z2}
\end{figure}

Fig.~\ref{fig:z2} shows the evolutions at $z=2$. The parameters of the
group are $R_\rmn{vir}=230$~kpc and $c_\rmn{vir}=3.3$. The shock reaches
the virial radius at $t\approx 6\times 10^7$~yr. The profile
$s_\rmn{f}/s_0$ in Fig.~\ref{fig:z2}(b) indicates that the ISM in the
galactic disc is stripped for $R<130$~kpc.

\begin{figure}
\includegraphics[width=74mm]{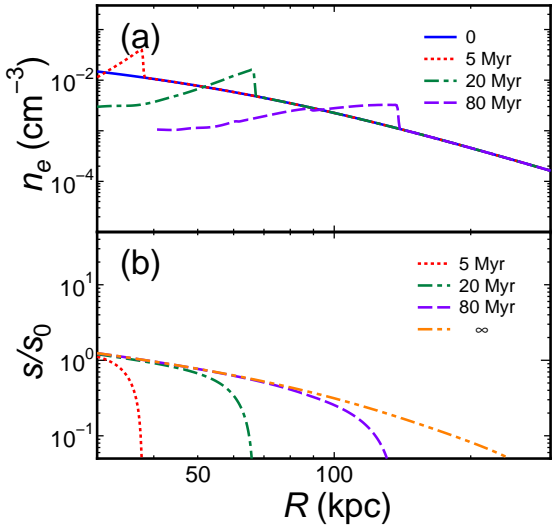} \caption{Same as Fig.~\ref{fig:z0}
 but for $z=2$ and $E_{\rm AGN}=6\times 10^{60}$~erg.}  \label{fig:z2_01}
\end{figure}

We also considered the case when the energy ejected by the AGN is
smaller. Fig.~\ref{fig:z2_01} shows the evolutions at $z=2$ when $E_{\rm
AGN}=6\times 10^{60}$~erg. The shock reaches the virial radius at
$t\approx 2\times 10^8$~yr. Since $s_\rmn{f}/s_0\la 1$ in the entire
group, the ISM in the galactic disc is not stripped
(Fig.~\ref{fig:z2_01}b).

\section{Discussion}
\label{sec:discuss}

We found that in the inner region of a group, the cold ISM in the disc
of a galaxy is stripped by the shock wave if $E_{\rm AGN}=6\times
10^{61}\rm\: erg\: s^{-1}$. We compare the effect with that of the usual
ram-pressure stripping through the motion of a galaxy in the IGM. The
typical velocity of galaxies in a group is
\begin{equation}
\label{eq:vgal}
 v_\rmn{gal}=\sqrt{G M_\rmn{vir}/R_\rmn{vir}}\:.
\end{equation}
For the group with $M_\rmn{vir}=10^{13}\rm\: M_\odot$, the velocity is
$v_\rmn{gal}=280\rm\: km\: s^{-1}$ at $z=0$ and $v_\rmn{gal}=440\rm\:
km\: s^{-1}$ at $z=2$.

The condition of usual ram-pressure stripping owing to the motion of a
galaxy is determined by the long-term (larger than the dynamical time of
the galaxy) balance between the ram-pressure from the IGM and the
gravity of the galaxy. Thus, the ram-pressure stripping is effective
when
\begin{eqnarray}
  \label{eq:strip}
  \rho_{\rm IGM}v_{\rm gal}^2 
 & >& 2\pi G \Sigma_{\star} \Sigma_{\rm ISM} \nonumber\\
 & =& v_{\rm rot}^2 r_\rmn{gal}^{-1} 
  \Sigma_{\rm ISM} \label{eq:grav2} \nonumber\\
 & =& 2.1\times 10^{-11}{\rm dyn\: cm^{-2}}
               \left(\frac{v_{\rm rot}}{220\rm\; km\: s^{-1}}\right)^2
               \nonumber\\
 &  &   \times \left[\frac{r_\rmn{gal}(z)}{10\rm\; kpc}\right]^{-1}
               \left[\frac{\Sigma_{\rm ISM}(z)}
               {8\times 10^{20} 
                   m_{\rm p}\;\rm cm^ {-2}}\right] \label{eq:grav3}\:
\end{eqnarray}
\citep{gun72,fuj99}.  From equations~(\ref{eq:Sig_ISM}) and
(\ref{eq:r_gal}), one finds that the right hand of the relation is
proportional to $H(z)^2$ for a given $v_\rmn{rot}$. Thus, from
equations~(\ref{eq:vgal}) and (\ref{eq:strip}), the ram-pressure
stripping is effective for $n_\rmn{e}>1.4\times 10^{-2}\rm\: cm^{-2}$ at
$z=0$ and $n_\rmn{e}>5.0\times 10^{-2}\rm\: cm^{-2}$ at
$z=2$. Figs.~\ref{fig:z0}(a) and \ref{fig:z2}(a) indicate that the
effect of the ram-pressure stripping can be ignored. This means that the
stripping by a shock wave overwhelms the usual ram-pressure stripping in
groups. However, in groups and clusters with larger $M_\rmn{vir}$, the
ram-pressure stripping should be more effective because the typical
velocity of galaxies, $v_\rmn{gal}$, is larger.

Even if $s_\rmn{f}/s_0<1$, the shock wave would affect the evolution of
a galaxy. It is often assumed that the cold ISM in the disc of a galaxy
is supplied from the warm gas in the halo. If most of the gas in the
halo is stripped, the ISM in the disc is gradually consumed through star
formation without supply. This is followed by gradual decrease of star
formation activity of the galaxy, which is often called `strangulation'
and is considered to be associated with the evolution of galaxies in
galaxy groups \citep*{lar80,bal00,bek02,fuj04a,kaw07}. From X-ray
observations, \citet{ped06} actually detected the halo gas in a massive
galaxy NGC~5746. However, the rotation velocity of the galaxy is very
large ($v_{\rm rot}=318\rm\: km\: s^{-1}$) and the result could not
apply to the galaxy we consider here ($v_{\rm rot}=220\rm\: km\:
s^{-1}$). Therefore, we use the results of the observations of NGC~5170
($v_{\rm rot}=247\rm\: km\: s^{-1}$) for which the halo gas was not
detected \citep{ped06,ras06}.

For our cosmological parameters, the upper limit of the mass of the halo
gas is $1.4\times 10^9\rm\: M_\odot$ \citep{ras06}. Assuming that the
halo gas resides in a cylinder of a base radius of 38~kpc \citep{ras06},
the column density in the direction of the height is $<4\times 10^{19}\:
m_{\rm p}\rm\: cm^{-2}$. Therefore, we assume that the column density of
the halo gas is $\Sigma_{\rm halo}(0)<4\times 10^{19}\: m_{\rm p}\rm\:
cm^{-2}$ and $\Sigma_{\rm halo}(z)\propto H(z)$ as equation
(\ref{eq:Sig_ISM}). Since $\Sigma_{\rm halo}v_{\rm esc}< 0.05\Sigma_{\rm
ISM}v_{\rm esc}=0.05\: s_0$ for our model galaxy, the halo gas would be
stripped even when $s_\rmn{f}/s_0\sim 0.05$ or smaller.

For $E_{\rm AGN}=6\times 10^{61}\rm\: erg$, Fig.~\ref{fig:z0}(b) and
Fig.~\ref{fig:z2}(b) show that $s_\rmn{f}/s_0\ga 0.05$ up to the virial
radius of the group (560~kpc for $z=0$ and 230~kpc for $z=2$). Thus,
most galaxies in the group would be affected by the outburst of the
central AGN. Even if $E_{\rm AGN}=6\times 10^{60}\rm\: erg$, the
influence of the AGN outburst cannot be ignored for $R<R_{\rm vir}$
(Fig.~\ref{fig:z2_01}b).

After the cold ISM (or the halo gas) is stripped by the outburst, the
colour of the galaxies would become red due to the lack of star
formation. Some of the galaxies would be observed as passive spiral
galaxies \citep[e.g.][]{got03}. Some would also fall into the group
centre through dynamical friction, and eventually merge with the central
galaxy in which the AGN responsible for the outburst resides. Since the
fallen galaxies no longer have ISM, they do not supply gas to the
supermassive black hole in the AGN. Thus, the outburst would serve as a
kind of feedback mechanism for the growth and activity of the black
hole.

At low redshift, radio and X-ray observations indicate that the
stripping by AGN outbursts may not be common. After the outburst, the
X-ray luminosities of the groups we considered reduce to $\sim 0.1$\% of
the initial values and become $\sim 10^{40}\:\rm erg\: s^{-1}$ for
$R<0.3\: R_{\rm vir}$. Here, we do not integrate the luminosities up to
$R_{\rm vir}$, because it is difficult to detect X-ray in the outermost
regions of groups \citep{mul00}. The luminosities of the IGM after the
outburst are smaller than those of groups from which X-ray emission from
the IGM has been detected ($\sim 10^{42}\rm\: erg\: s^{-1}$;
\citealt{mul00}). If the stripping by AGN outbursts were common, the
lack of cold ISM in galaxies should be confirmed more in groups lacking
for X-ray emission. However, radio (H{\footnotesize I}) and X-ray
observations of nearby groups showed the opposite trend
\citep*{sen07,ver07}, although the cause is not known.

On the other hand, we suppose that the stripping by AGN outbursts
occurred more often at high redshift (say $z\sim 2$). This is because
strong outbursts were more common at that time. For example, radio
observations showed that there are a number of AGNs with the jet power
of $>10^{46}\rm\: erg\: s^{-1}$ at $z\ga 0.5$
\citep{raw91,dal95}. Considering typical duration of an AGN activity
($\sim 10^7$--$10^8$~yr; \citealt{raw91}), the total energy injected
through an activity is $\ga 10^{61}$~erg. Moreover, it has been
indicated that luminous AGNs tend to be found at higher redshifts
($z\sim 2$) in comparison with less luminous AGNs
\citep{ued03,kau07}. Furthermore, optical observations often found
clustering of red galaxies around radio galaxies at $z\ga 1$
\citep[e.g.][]{nak01,kaj06}. The activities of the radio galaxies might
have affected the star formation of the surrounding galaxies as we
predict. From a theoretical point of view, \citet{fuj01} indicated that
strong AGN outbursts at $z\ga 1$ blew out the IGM from groups (ancestors
of present-day groups or clusters), and they could be responsible for
the low metal abundance observed in groups ($\sim 10^{14}\rm\: M_\odot$)
at $z\sim 0$ \citep{ren97}. Although these studies do not directly prove
the stripping by AGN outbursts, they are at least consistent with our
model. In the future, galaxies lack of cold gas would be observed inside
the cocoon or the shock produced by an outburst.

\section{Conclusions}

We have investigated the influence of an AGN outburst at the centre of a
galaxy group on the galaxies surrounding the AGN. If the energy of the
AGN is extremely large ($E_{\rm AGN}\sim 6\times 10^{61}\rm\; erg$) as
the one recently observed in clusters, the shock wave excited by the
outburst strips the cold ISM in the disc of the surrounding galaxies
locating at $R\sim 100$--200~kpc from the AGN. The effect of the
stripping can be much stronger than that of usual ram-pressure stripping
owing to the motion of the galaxies in a group. However, the stripping
of the cold ISM in the disc is not effective if $E_{\rm AGN}\sim 6\times
10^{60}\rm\; erg$.

We also showed that even if the cold ISM in the disc of the galaxies is
not stripped, the warm gas in the halo would be stripped by the shock
wave. This could be effective for the whole galaxies in the group up to
the virial radius even when $E_{\rm AGN}\sim 6\times 10^{60}\rm\;
erg$. The star formation activities of those galaxies would decrease
gradually because the ISM consumed by the star formation is not
compensated from the gas in the halo (strangulation).  Our results
indicate that even one strong AGN outburst significantly affects the
evolution of galaxies in groups. After the stripping, the galaxies
should become red because of the decrease of the star formation rate.

The success of this model depends on the assumption of an extremely
strong outburst. Since such outbursts have not been observed in galaxy
groups at low-redshift, the stripping by them would not have strong
influence on the evolution of galaxies at low-redshift. However, we
suppose that it would be at high redshift where extremely strong AGN
activities have often been observed.

\section*{Acknowledgments}

The author wishes to thank the referee for useful comments. Y. F. was
supported in part by Grants-in-Aid from the Ministry of Education,
Science, Sports, and Culture of Japan (17740162).

\label{lastpage}


\begin{thebibliography}{99}
%
\bibitem[\protect\citeauthoryear{Allen \& 
Fabian}{1998}]{all98} Allen S.~W., Fabian A.~C., 1998, MNRAS, 
297, L57 

\bibitem[\protect\citeauthoryear{Arnaud \& 
Evrard}{1999}]{arn99} Arnaud M., Evrard A.~E., 1999, MNRAS, 
305, 631 

\bibitem[\protect\citeauthoryear{Balogh, Navarro, \& 
Morris}{Balogh et al.}{2000}]{bal00} Balogh M.~L., Navarro J.~F., Morris
S.~L., 2000, ApJ, 540, 113 

\bibitem[\protect\citeauthoryear{Bekki, Couch, \& 
Shioya}{Bekki et al.}{2002}]{bek02} Bekki K., Couch W.~J., Shioya Y.,
	      2002,
ApJ, 577, 651 

\bibitem[\protect\citeauthoryear{Blanton et 
al.}{2001}]{bla01} Blanton E.~L., Sarazin C.~L., McNamara 
B.~R., Wise M.~W., 2001, ApJ, 558, L15 

\bibitem[\protect\citeauthoryear{Bower et al.}{2001}]{bow01} 
Bower R.~G., Benson A.~J., Lacey C.~G., Baugh C.~M., Cole S., Frenk C.~S., 
2001, MNRAS, 325, 497 

\bibitem[\protect\citeauthoryear{Bullock et 
al.}{2001}]{bul01} Bullock J.~S., Kolatt T.~S., Sigad Y., 
Somerville R.~S., Kravtsov A.~V., Klypin A.~A., Primack J.~R., Dekel A., 
2001, MNRAS, 321, 559 

\bibitem[\protect\citeauthoryear{Bryan \& 
Norman}{1998}]{bry98} Bryan G.~L., Norman M.~L., 1998, ApJ, 
495, 80 

\bibitem[\protect\citeauthoryear{Daly}{1995}]{dal95} Daly 
R.~A., 1995, ApJ, 454, 580 

\bibitem[\protect\citeauthoryear{Edge \& 
Stewart}{1991}]{edg91} Edge A.~C., Stewart G.~C., 1991, 
MNRAS, 252, 414 

\bibitem[\protect\citeauthoryear{Fabian et al.}{2000}]{fab00} 
Fabian A.~C., et al., 2000, MNRAS, 318, L65 

\bibitem[\protect\citeauthoryear{Fujita}{2001}]{fuj01} Fujita 
Y., 2001, ApJ, 550, L7 

\bibitem[\protect\citeauthoryear{Fujita}{2004}]{fuj04a} Fujita 
Y., 2004, PASJ, 56, 29 

\bibitem[\protect\citeauthoryear{Fujita \& 
Goto}{2004}]{fuj04} Fujita Y., Goto T., 2004, PASJ, 56,
	    621 

\bibitem[\protect\citeauthoryear{Fujita \& 
Nagashima}{1999}]{fuj99} Fujita Y., Nagashima M., 1999, ApJ, 
516, 619 

\bibitem[\protect\citeauthoryear{Fujita et al.}{2007}]{fuj07} 
Fujita Y., Kohri K., Yamazaki R., Kino M., 2007, ApJ, 663, L61

\bibitem[\protect\citeauthoryear{Fukushige \& 
Makino}{1997}]{fuk07} Fukushige T., Makino J., 1997, ApJ, 
477, L9 

\bibitem[\protect\citeauthoryear{Gitti et al.}{2007}]{git07} 
Gitti M., McNamara B.~R., Nulsen P.~E.~J., Wise M.~W., 2007, ApJ, 660,
		1118 

\bibitem[\protect\citeauthoryear{Goto et al.}{2003}]{got03} 
Goto T., et al., 2003, PASJ, 55, 757 

\bibitem[\protect\citeauthoryear{Gunn \& Gott}{1972}]{gun72} 
Gunn J.~E., Gott J.~R.~I., 1972, ApJ, 176, 1 

\bibitem[\protect\citeauthoryear{Inoue \& 
Sasaki}{2001}]{ino01} Inoue S., Sasaki S., 2001, ApJ, 562, 
618 

\bibitem[\protect\citeauthoryear{Kaiser}{1986}]{kai86} Kaiser 
N., 1986, MNRAS, 222, 323 

\bibitem[\protect\citeauthoryear{Kajisawa et 
al.}{2006}]{kaj06} Kajisawa M., Kodama T., Tanaka I., Yamada 
T., Bower R., 2006, MNRAS, 371, 577 

\bibitem[\protect\citeauthoryear{Kauffmann et al.}{2007}]{kau07}
Kauffmann, G., Heckman, T.~M., Best, P.~N. 2007, arXiv:0709.2911,
submitted to MNRAS
		 

\bibitem[\protect\citeauthoryear{Kawata \& Mulchaey}{2007}]{kaw07} 
Kawata, D., Mulchaey, J.~S. 2007, arXiv:0707.3814, submitted to ApJL

\bibitem[\protect\citeauthoryear{Komatsu \& 
Seljak}{2001}]{kom01} Komatsu E., Seljak U., 2001, MNRAS, 
327, 1353 

\bibitem[\protect\citeauthoryear{Larson, Tinsley, \& 
Caldwell}{Larson et al.}{1980}]{lar80} Larson R.~B., Tinsley B.~M.,
	      Caldwell
C.~N., 1980, ApJ, 237, 692 

\bibitem[\protect\citeauthoryear{Loewenstein}{2000}]{loe00} 
Loewenstein M., 2000, ApJ, 532, 17 

\bibitem[\protect\citeauthoryear{McNamara et 
al.}{2000}]{mcn00} McNamara B.~R., et al., 2000, ApJ, 534, 
L135 

\bibitem[\protect\citeauthoryear{McNamara et 
al.}{2005}]{mcn05} McNamara B.~R., Nulsen P.~E.~J., Wise 
M.~W., Rafferty D.~A., Carilli C., Sarazin C.~L., Blanton E.~L., 2005, 
Nature, 433, 45 

\bibitem[\protect\citeauthoryear{Markevitch}{1998}]{mar98} 
Markevitch M., 1998, ApJ, 504, 27 

\bibitem[\protect\citeauthoryear{Menci \& 
Cavaliere}{2000}]{men00} Menci N., Cavaliere A., 2000, MNRAS, 
311, 50 

\bibitem[\protect\citeauthoryear{Mo, Mao, \& 
White}{Mo et al.}{1998}]{mo98} Mo H.~J., Mao S., White S.~D.~M., 1998, 
MNRAS, 295, 319

\bibitem[\protect\citeauthoryear{Mulchaey}{2000}]{mul00} 
Mulchaey J.~S., 2000, ARA\&A, 38, 289 

\bibitem[\protect\citeauthoryear{Nakata et al.}{2001}]{nak01} 
Nakata F., et al., 2001, PASJ, 53, 1139 

\bibitem[\protect\citeauthoryear{Navarro, Frenk, \& 
White}{Navarro et al.}{1997}]{nav97} Navarro J.~F., Frenk C.~S., White
S.~D.~M., 1997, ApJ, 490, 493 

\bibitem[\protect\citeauthoryear{Nulsen et al.}{2005a}]{nul05a} 
Nulsen P.~E.~J., Hambrick D.~C., McNamara B.~R., Rafferty D., Birzan L., 
Wise M.~W., David L.~P., 2005a, ApJ, 625, L9 

\bibitem[\protect\citeauthoryear{Nulsen et al.}{2005b}]{nul05b} 
Nulsen P.~E.~J., McNamara B.~R., Wise M.~W., David L.~P., 2005b, ApJ,
	      628,
629 

\bibitem[\protect\citeauthoryear{Pedersen et 
al.}{2006}]{ped06} Pedersen K., Rasmussen J., Sommer-Larsen 
J., Toft S., Benson A.~J., Bower R.~G., 2006, NewA, 11, 465 

\bibitem[\protect\citeauthoryear{Ponman, Cannon, \& 
Navarro}{Ponman et al.}{1999}]{pon99} Ponman T.~J., Cannon D.~B.,
	      Navarro
J.~F., 1999, Nature, 397, 135 

\bibitem[\protect\citeauthoryear{Rasmussen et 
al.}{2006}]{ras06} Rasmussen J., Sommer-Larsen J., Pedersen 
K., Toft S., Benson A., Bower R.~G., Olsen L.~F., 2006, 
arXiv:astro-ph/0610893, submitted to ApJ

\bibitem[\protect\citeauthoryear{Rawlings \& 
Jarvis}{2004}]{raw04} Rawlings S., Jarvis M.~J., 2004, MNRAS, 
355, L9 

\bibitem[\protect\citeauthoryear{Rawlings \& 
Saunders}{1991}]{raw91} Rawlings S., Saunders R., 1991, 
Nature, 349, 138 

\bibitem[\protect\citeauthoryear{Renzini}{1997}]{ren97} 
Renzini A., 1997, ApJ, 488, 35 

\bibitem[\protect\citeauthoryear{Sengupta, Balasubramanyam, \& 
Dwarakanath}{Sengupta et al.}{2007}]{sen07} Sengupta C., Balasubramanyam
	      R.,
Dwarakanath K.~S., 2007, MNRAS, 378, 137 

\bibitem[\protect\citeauthoryear{Spitzer}{1978}]{spi78} Spitzer,
	   L. 1978, Physical Processes in
			      the Interstellar Medium (New York: Wiley)

\bibitem[\protect\citeauthoryear{Ueda et al.}{2003}]{ued03} 
Ueda Y., Akiyama M., Ohta K., Miyaji T., 2003, ApJ, 598, 886 

\bibitem[\protect\citeauthoryear{Valageas \& 
Silk}{1999}]{val99} Valageas P., Silk J., 1999, A\&A, 350, 
725 

\bibitem[\protect\citeauthoryear{Verdes-Montenegro et 
al.}{2007}]{ver07} Verdes-Montenegro L., Yun M.~S., Borthakur 
S., Rasmussen J., Ponman T., 2007, NewAR, 51, 87 

\bibitem[\protect\citeauthoryear{Wise et al.}{2007}]{wis07} 
Wise M.~W., McNamara B.~R., Nulsen P.~E.~J., Houck J.~C., David L.~P., 
2007, ApJ, 659, 1153 

\bibitem[\protect\citeauthoryear{Wu, Fabian, \& 
Nulsen}{Wu et al.}{1998}]{wu98} Wu K.~K.~S., Fabian A.~C., Nulsen 
P.~E.~J., 1998, MNRAS, 301, L20 

\bibitem[\protect\citeauthoryear{Wu, Fabian, \& 
Nulsen}{Wu et al.}{2000}]{wu00} Wu K.~K.~S., Fabian A.~C., Nulsen 
P.~E.~J., 2000, MNRAS, 318, 889 

\end{thebibliography}
\end{document}